\documentclass[fleqn]{article}
\usepackage{amsmath,amssymb}
\usepackage{epsfig,graphicx}

\renewcommand{\theequation}{\arabic{section}.\arabic{equation}}
\renewcommand{\thesection}{\arabic{section}.}

\catcode`@=11 \@addtoreset{equation}{section} \catcode`@=12

\begin{document}
\title{\vskip-1.7cm \bf  The path integral for the statistical
sum of the microcanonical ensemble in cosmology}
\date{}
\author{A.O.Barvinsky$^{1,\,2}$}
\maketitle \hspace{-8mm} {\,\,$^{1}$\em
Theory Department, Lebedev
Physics Institute, Leninsky Prospect 53, Moscow 119991, Russia\\
$^{2}$Department of Physics, Ludwig Maximilians University,
Theresienstrasse 37, Munich, Germany}
\begin{abstract}
The path integral is calculated for the statistical sum of the
microcanonical ensemble in a generic time-parametrization invariant
gravitational model with the Friedman-Robertson-Walker (FRW) metric.
This represents the first example of a systematic calculation of the
Faddeev-Popov gauge-fixed path integral in the minisuperspace sector
of quantum cosmology. The gauge fixing procedure, together with
gauging out local diffeomorphisms, also handles the residual
symmetries associated with the conformal Killing vector of the FRW
metric and incorporates the Batalin-Vilkovisky quantization
technique for gauge theories with linearly dependent generators. For
a subset of saddle-point instantons, characterized by a single
oscillation of the FRW scale factor, this technique is designed to
obtain the one-loop statistical sum in the recently suggested model
of cosmological initial conditions generated by a conformal field
theory with a large number of quantum species.
\end{abstract}

\section{Introduction}
The purpose of this paper is to calculate beyond tree-level
approximation the path integral for the statistical sum in the
generic time-parametrization invariant gravitational system with the
Friedman-Robertson-Walker (FRW) metric. This problem is motivated by
the recently suggested model of initial conditions in cosmology in
the form of the microcanonical density matrix \cite{slih,why}. In
contrast to the formal Euclidean quantum gravity origin of the
no-boundary prescription \cite{noboundary} or essentially
semiclassical tunneling prescription \cite{tunnel0} for the
cosmological state, this model has a clear origin in terms of
operator quantization of gravity theory in the Lorentzian signature
spacetime. In particular, it is based on a natural notion of the
microcanonical density matrix as a projector on the space of
solutions of the quantum gravitational Dirac constraints -- the
system of Wheeler-DeWitt equations \cite{why}. Moreover, when
applied to cosmology with a large number of fields conformally
coupled to gravity, this theory can be important within the
cosmological constant and dark energy problems. In particular, its
statistical ensemble is bounded to a finite range of values of the
effective cosmological constant, it generates an inflationary stage
and is potentially capable of generating the cosmological
acceleration phenomenon within the so-called Big Boost scenario
\cite{bigboost}.

As shown in \cite{why}, for a spatially closed cosmology with
$S^3$-topology the microcanonical statistical sum can be represented
by the Euclidean quantum gravity path integral,
    \begin{eqnarray}
    &&Z=
    \!\!\int\limits_{\,\,\rm periodic}
    \!\!\!\! D[\,g_{\mu\nu},\phi\,]\;
    e^{-S[\,g_{\mu\nu},\phi\,]},
    \end{eqnarray}
over the metric $g_{\mu\nu}$ and matter fields $\phi$ which are
periodic on the Euclidean spacetime with a compactified time $\tau$
(of $S^1\times S^3$ topology). The FRW metric arises in this path
integral as the set of major collective variables of cosmology.
Under the decomposition of the full set of $g_{\mu\nu}(x),\phi(x)$
into the minisuperspace FRW sector
    \begin{eqnarray}
    ds^2 =N^2(\tau)\,d\tau^2
    +a^2(\tau)\,d^2\Omega^{(3)}, \label{FRW}
    \end{eqnarray}
and inhomogeneous ``matter'' fields
$\varPhi(x)=(\phi(x),\psi(x),A_\mu(x), h_{\mu\nu}(x),...)$ on the
background of this metric the path integral can be cast into the
form of an integral over a minisuperspace lapse function $N(\tau)$
and a scale factor $a(\tau)$,
    \begin{eqnarray}
    &&Z=\int
     D[\,a,N\,]\;
    e^{-\varGamma[\,a,\,N\,]},              \label{1}\\
    &&e^{-\varGamma[\,a,\,N]}
    =\int D\varPhi(x)\,
    e^{-S[\,a,\,N;\,\varPhi(x)\,]}\ .             \label{2}
    \end{eqnarray}
Here, $\varGamma[\,a,\,N\,]$ is the Euclidean effective action of
the fields $\varPhi$ (which include also the metric perturbations
$h_{\mu\nu}$) on the FRW background, and
$S[\,a,N;\varPhi(x)\,]\equiv S_E[\,g_{\mu\nu},\phi\,]$ is the
original action rewritten in terms of this minisuperspace
decomposition. It is important that this representation is not a
minisuperspace approximation, when all the fields $\varPhi(x)$ are
frozen out. Rather this is the disentangling of the collective
degrees of freedom from the configuration space, the rest of which
effectively manifests itself in terms of this effective action.

In the theory with a primordial cosmological constant and a large
number of free (linear) fields conformally coupled to gravity --
conformal field theory (CFT) --  such effective action is dominated
by the contribution of these fields because they simply outnumber
the non-conformal fields including, in particular, the graviton.
Then this quantum effective action is exactly calculable as a {\em
functional} of histories $(a(\tau),N(\tau))$ by the conformal
transformation converting (\ref{FRW}) into the static (Einstein
Universe) metric with $a={\rm const}$ \cite{FHH,Starobinsky,conf}.
The structure of the resulting  action \cite{slih} is
    \begin{eqnarray}
    &&\varGamma[\,a,N\,]=
    \oint d\tau\,N {\cal L}(a,a')+ F(\eta),    \label{action1}\\
    &&\eta=\oint d\tau\,\frac{N}a.                   \label{eta}
    \end{eqnarray}
where $a'\equiv da/Nd\tau$ and the integration runs over the period
of $\tau$ on the circle $S^1$ of $S^1\times S^3$. Here the effective
Lagrangian of its local part ${\cal L}(a,a')$ includes the classical
Einstein term and the polarization effects of quantum fields and
their vacuum (Casimir) energy. A nonlocal part of the action
$F(\eta)$ is the free energy of their quasi-equilibrium excitations
with the temperature given by the inverse of the conformal time
(\ref{eta}). This is a typical boson or fermion sum
$F(\eta)=\pm\sum_{\omega}\ln\big(1\mp e^{-\omega\eta}\big)$ over
field oscillators with energies $\omega$ on a unit 3-sphere. In what
follows the concrete structure of ${\cal L}(a,a')$ and $F(\eta)$
will be unimportant for us -- remarkably the one-loop functional
integration can be done in a closed form without specifying a
concrete form of these functions. The only restriction will be the
absence of higher derivatives of $a$ in ${\cal L}(a,a')$, which of
course implies a special UV renormalization of the effective action
(\ref{2}). Such a renormalization really exists, as was shown for
the CFT driven cosmology \cite{slih}, and it does not introduce into
the minisuperspace sector of Einstein theory extra degrees of
freedom.\footnote{This choice of renormalization was motivated in
\cite{slih} by ghost-free requirements and certain universality
properties which, in particular, relate the value of the Casimir
energy to the coefficient the conformal anomaly of CFT fields -- the
source of ${\cal L}(a,a')$ in (\ref{action1})
\cite{universality,DGP/CFT}.}

Semiclassically the integral (\ref{1}) is dominated by the saddle
points --- solutions of the effective Friedmann equation of the {\em
Euclidean} gravity theory, which for a generic Lagrangian of the
above type reads as
    \begin{eqnarray}
    &&\frac{\delta\varGamma}{\delta N}
    ={\cal L}-\frac{\partial\cal
    L}{\partial a'}\,a'+\frac1a\,\frac{dF}{d\eta}=0, \label{efeq}
    \end{eqnarray}
The term with $dF/d\eta$ here characterizes the energy of the gas of
thermally excited particles with the inverse temperature $\eta$ --
the instanton period in units of the conformal time. The latter is
given by the integral (\ref{eta}) over the full period of $\tau$ or
the $2k$-multiple of the integral between the two neighboring
turning points of the scale factor history $a(\tau)$,
$a'(\tau_\pm)=0$.

This $k$-fold nature follows from the fact that in a periodic
solution of (\ref{efeq}) the scale factor necessarily oscillates
certain number of times between its maximum and minimum values
$a_\pm=a(\tau_\pm)$, $a_-\leq a(\tau)\leq a_+$, and forms a kind of
a garland of $S^1\times S^3$ topology with oscillating $S^3$
sections. These garland-type instantons  are weighted in the
relevant statistical ensemble by their exponentiated on-shell action
    \begin{eqnarray}
    &&Z=P\,\exp\big(-\varGamma_0),
    \end{eqnarray}
where $\varGamma_0=\varGamma[\,a,N\,]$ is taken at the solution of
Eqs.(\ref{eta})-(\ref{efeq}) and $P$ is the preexponential factor
accumulating quantum corrections of the semiclassical expansion. For
a particular case of the CFT driven cosmology \cite{slih}, these
instantons exist only in the limited range of the cosmological
constant $\Lambda=3H^2$, $0<H^2_{\rm min}<H^2< H^2_{\rm max}$ and,
thus, might be relevant to the solution of the cosmological constant
problem \cite{slih,why}. Here we will focus on the calculation of
the one-loop prefactor $P$ for a generic action (\ref{action1}).
Application of this calculation to the CFT driven cosmology will be
considered in a sequel to this paper \cite{oneloop}.

In fact, this represents the first example of a systematic and
explicit one-loop calculation of the gauge-fixed path integral in
the minisuperspace sector of quantum cosmology. Compared to early
examples of the path integral calculation, that were basically
focused on the formal derivation of the Wheeler-DeWitt equations
\cite{Barv86,Halliwell88,HalHartle,BarvU} or applications to
cosmological toy models \cite{GHMM91}, our results here apply to the
models with the effective minisuperspace action of a rather generic
form (\ref{action1}) which includes according to (\ref{2}) the
effect of inhomogeneous field modes. Therefore, this is not a
minisuperspace approximation freezing out all local inhomogeneous
degrees of freedom. Moreover, our calculations of the statistical
sum differ by boundary conditions -- periodic boundary conditions vs
the Dirichlet (or mixed Neumann-Dirichlet \cite{NeumannDirichlet})
boundary conditions for the unitary transition amplitude of
\cite{Halliwell88,HalHartle,BarvU,MukhanovAnderegg}. This leads to
an additional peculiarity of the formalism which finally amounts to
a special gauge fixing procedure -- the one for the system with
linearly dependent generators of gauge transformations -- and, thus,
goes beyond conventional Faddeev-Popov technique.

The nature of this peculiarity is as follows. By counting the number
of local degrees of freedom the minisuperspace sector of $a$ and $N$
is not dynamical (a second order action with two-dimensional
configuration space and one local gauge invariance), and the
one-loop prefactor $P$ seems being a trivial normalization constant.
This conclusion is, however, misleading because of the global degree
of freedom related to the periodic nature of $S^1\times S^3$ and its
interplay with the zero mode associated with the conformal Killing
symmetry of a generic FRW metric (\ref{FRW}). Gauging this symmetry
out effectively leads to the model with linearly dependent
generators, which requires the use of the Batalin-Vilkovisky
quantization technique \cite{BFV}. Application of this technique
shows that the cancelation in the preexponential factor, associated
with the absence of local degrees of freedom, has only a partial
nature and leaves us with a nontrivial contribution to $P$ of the
CFT radiation bath (namely its specific heat $d^2F/d\eta^2$).

Below we calculate this contribution for a subset of background
instantons, corresponding to the solutions of (\ref{efeq}) with a
single oscillation of the scale factor $a(\tau)$. After a brief
formulation of main results in Sect.2, we begin this one-loop
calculation in Sect.3 with the derivation of the quadratic part of
the action, which turns out to be parameterized by a single function
$g(\tau)$ -- the zero mode of the operator of small disturbances on
the instanton background. Then in Sect.4 we describe the gauge
fixing procedure which reveals the residual conformal Killing
invariance of the action and incorporates the set of linearly
dependent generators. The necessary Batalin-Vilkovisky technique
\cite{BFV} is then briefly presented in Sect.5. In Sects.6 and 7 we
calculate the contributions of the gauge and ghost sectors of the
path integral and the sector of metric perturbations, which lead to
the final closed algorithm for $P$. We accomplish the paper with
concluding remarks in Conclusions. Two appendices contain the
derivation of gauge independence properties in the
Batalin-Vilkovisky technique for systems with linearly dependent
generators and the treatment of a redundant set of gauge conditions
by means of the so-called extraghost \cite{BFV}.

\section{The formulation of main results}
The action (\ref{action1})-(\ref{eta}) with the Lagrangian ${\cal
L}(a,a')$ -- a rather generic function of $a$ and its
parametrization invariant derivative $a'\equiv da/Nd\tau$ -- is
invariant with respect to local reparametrizations of time,
$\tau\to\bar\tau=\bar\tau(\tau)$,
    \begin{eqnarray}
    \bar a(\bar\tau)= a(\tau),\,\,\,\,
    \bar N(\bar\tau)=
    \left(\frac{d\bar\tau}{d\tau}\right)^{-1}N(\tau).   \label{diffeo}
    \end{eqnarray}
As we show below, the gauge-fixed path integral (\ref{1}) for the
statistical sum of this model can be written down in the one-loop
approximation in terms of a special set of variables $\varphi$ and
$n$ as
    \begin{eqnarray}
    &&Z=e^{-\varGamma_0}\int Dn\,
    \delta[\,n'(\tau)\,]\;\big(\,{\rm
    Det_*}\,{\mbox{\boldmath$Q$}}\,\big)\nonumber\\
    &&\qquad\qquad\qquad\qquad\quad
    \times\int D\varphi\,
    \delta\left(\oint d\tau\,g\varphi\right)\,Q\,
    \exp\Big(-\varGamma_{(2)}
    [\,\varphi,n\,]\Big).                    \label{statsum1}
    \end{eqnarray}
Here the delta function of the gauge condition $n'$ and the relevant
Faddeev-Popov functional determinant ${\rm
Det_*}\,{\mbox{\boldmath$Q$}}$ gauge out the local
time-parametrization invariance of the action
$\varGamma_{(2)}[\,\varphi,n\,]$ quadratic in its arguments. The
star denotes the removal of the residual gauge invariance with
respect to the conformal Killing transformations: the symbol of the
{\em restricted} functional determinant ${\rm
Det_*}\,{\mbox{\boldmath$Q$}}$ implies the omission of the relevant
zero mode of ${\mbox{\boldmath$Q$}}$. This residual gauge
transformation is the invariance of $\varGamma_{(2)}[\,\varphi,n\,]$
under the global transformation
$\varphi(\tau)\to\varphi(\tau)+\Delta^\varepsilon\varphi(\tau)$,
$\Delta^\varepsilon\varphi(\tau)\equiv\varepsilon\,g(\tau)$, with
the function $g(\tau)$ -- the zero mode of the operator of field
disturbances $\varphi$,
    \begin{eqnarray}
    &&{\mbox{\boldmath$F$}}
    =\frac{\delta^2\varGamma_{(2)}}
    {\delta\varphi(\tau)\,
    \delta\varphi(\tau')},               \label{F0}\\
    &&{\mbox{\boldmath$F$}}g=0.
    \end{eqnarray}
It is gauged out in the integral over $\varphi$ by the delta
function of the (time-nonlocal) gauge condition
$\chi[\,\varphi\,]=\oint d\tau\,g\varphi$, which implies a
functional orthogonality of $\varphi(\tau)$ to the zero mode
$g(\tau)$. This is accompanied by the Faddeev-Popov factor $Q$,
$\Delta^\varepsilon\chi\equiv Q\varepsilon$,
    \begin{eqnarray}
    Q=\oint d\tau\,g^2(\tau).      \label{FPzeromode}
    \end{eqnarray}

The Gaussian integration over $\varphi$ and the integration over
$n(\tau)$, which in view of the gauge condition $n'=0$ reduces to
the integral over the constant mode $n_0={\rm const}$, give the
preexponential factor of (\ref{statsum1})
    \begin{eqnarray}
    &&P={\rm const}\times\big(\,{\rm Det_*}\,
    {\mbox{\boldmath$F$}}\,\big)^{-1/2}\int dn_0\,
    \exp\left(\,n_0^2\,
    \frac{Y}{\mbox{\boldmath$I$}}\right),     \label{Zoneloop}
    \end{eqnarray}
where the coefficient $Y$ differs from unity by the contribution of
the specific heat $d^2F(\eta)/d\eta^2$ of conformal modes, whereas
the restricted functional determinant ${\rm
Det_*}\,{\mbox{\boldmath$F$}}$ (with its zero mode gauged out) is
given exactly by the factor ${\mbox{\boldmath$I$}}$ which is
contained above in the denominator of the exponential -- the
property proven in the accompanying paper \cite{det},
    \begin{eqnarray}
    &&Y=1-{\mbox{\boldmath$I$}}\,\frac{d^2F}{d\eta^2},   \label{Y}\\
    &&{\rm Det_*}\,{\mbox{\boldmath$F$}}
    ={\rm const}\times {\mbox{\boldmath$I$}}.   \label{DetFstar}
    \end{eqnarray}
Therefore, after Gaussian integration over $n_0$ the resulting
one-loop prefactor $P$ differs from a trivial normalization constant
only by the contribution of matter sector $O(d^2F/d\eta^2)$ to $Y$
and reads
    \begin{eqnarray}
    P=\frac{\rm const}{\sqrt{\,|\,Y|\,}}.    \label{prefactor0}
    \end{eqnarray}

The details of the above mechanism are as follows. To begin with,
the new variables, which parameterize the perturbations of the scale
factor $\delta a$ and the lapse function $\delta N$ on the
background of the solution of the equation of motion for
(\ref{action1}), read as
    \begin{eqnarray}
    &&\delta a=\frac{aa'}g\,\varphi,     \label{newvariables1}\\
    &&\delta N=\frac{a'}g\,\varphi+na.    \label{newvariables2}
    \end{eqnarray}
Here the function $g=g(\tau)$ -- the zero mode introduced above --
expresses in terms of the Hessian of the Lagrangian with respect to
the scale factor ``velocity" $a'$,
    \begin{eqnarray}
    g=a'a\sqrt{|{\cal D}|},\,\,\,\,
    {\cal D}=\frac{\partial^2\cal
    L}{\partial a'\partial a'}.            \label{g}
    \end{eqnarray}
The most important property of these variables is that $\varphi$ is
canonically normalized, and that the local part of their quadratic
action has a simple closed form which is universally parametrized by
the same single function $g(\tau)$,
    \begin{eqnarray}
    &&\varGamma_{(2)}[\,\varphi,n\,]=
    \frac12\,\varepsilon_{\cal D}\oint
    d\tau\,\left\{\varphi'^2+\frac{g''}g\,\varphi^2
    +\big(\,4ng'+2n'g\big)\varphi+g^2n^2\right\}\nonumber\\
    &&\qquad\qquad\qquad\quad+
    \frac12\,\frac{d^2F}{d\eta^2}
    \left(\oint d\tau\, n\right)^2            \label{quadraction}\\
    &&\varepsilon_{\cal D}
    =\frac{\cal D}{|\,\cal D\,|}=\pm1.
    \end{eqnarray}
This property holds irrespective of the form of the Lagrangian
${\cal L}(a,a')$, and the source of this universality is, of course,
the time-parametrization invariance of the action and the fact that
${\cal L}(a,a')$ does not contain higher order derivatives of $a$.

The operator (\ref{F0}) for this action equals
    \begin{eqnarray}
    {\mbox{\boldmath$F$}}=
    -\frac{d^2}{d\tau^2}+\frac{g''}g.             \label{operator}
    \end{eqnarray}
It has as the zero mode the periodic regular function $g(\tau)$.
Other important properties of this operator follow from the
following observations.

The one-fold instanton solution (which only we consider in this
paper) has one oscillation of the scale factor between its maximal
and minimal values $a_\pm=a(\tau_\pm)$. Therefore, the function
$g(\tau)\sim a'(\tau)$ has two zeroes at these points,
$g(\tau_\pm)=0$, which mark the boundaries of the half period of the
total time range, $T=2(\tau_+-\tau_-)$. For brevity of the formalism
we shift the point of the minimal $a_-$ to zero, $\tau_-=0$, and let
the coordinate $\tau$ run in the total range
$-\tau_+\leq\tau\leq\tau_+$ with the points $\pm\tau_+$ identified.
Then $g(\tau)$ is an odd function of $\tau$ which is periodic with
all its derivatives and has two first degree zeros at antipodal
points $\tau=\tau_-\equiv 0$ and $\tau=\tau_+$ of this circle
    \begin{eqnarray}
    &&g(\tau)=-g(-\tau),\\
    &&g(\tau_\pm)=0,\,\,\,g'(\tau_\pm)
    \equiv g'_\pm\neq 0.                    \label{conditionsong}
    \end{eqnarray}
In spite of singularity of $1/g$ at $\tau_\pm$ the operator
(\ref{operator}) is everywhere regular (analytic) on the circle,
because from the equation of motion (\ref{efeq}) it follows that all
odd order derivatives of $a$ at $\tau_\pm$ vanish, and
$g''(\tau_\pm)=0$ (this is guaranteed by the assumption that ${\cal
L}(a,a')$ is an even function of $a'$).

The final result of this paper -- the prefactor
(\ref{Y})-(\ref{prefactor0}) -- is determined by the second solution
of the homogeneous equation ${\mbox{\boldmath$F$}}\,\varPsi=0$,
which together with $g(\tau)$ forms a full set of basis functions of
${\mbox{\boldmath$F$}}$. This is a two-point function
$\Psi(\tau,\tau_*)$
    \begin{eqnarray}
    \varPsi(\tau,\tau_*)\equiv
    g(\tau)\int_{\tau_*}^{\tau}\frac{dy}{g^2(y)},\,\,\,\,\,
    \tau_-\equiv 0<\tau<\tau_+,\,\,\,\,\,
    \tau_-<\tau_*<\tau_+,                              \label{Psi}
    \end{eqnarray}
with some fixed point $\tau_*$ in the half period range of the
instanton time. This function is smoothly defined only in the
half-period range of $\tau$ and $\tau_*$, because otherwise the
integral for $\varPsi(\tau,\tau_*)$ is divergent if the roots of
$g(\tau)$ lie between $\tau$ and $\tau_*$. Therefore (\ref{Psi})
cannot be smoothly continued beyond the half-period
$\tau_-\leq\tau\leq\tau_+$, though its limits are well defined for
$\tau\to\tau_\pm\mp 0$,
    \begin{eqnarray}
    \varPsi(\tau_\pm,\tau_*)=
    -\frac1{g'(\tau_\pm)}\equiv
    -\frac1{g'_\pm},                  \label{Psipm}
    \end{eqnarray}
because the factor $g(\tau)$ tending to zero compensates for the
divergence of the integral at $\tau\to\tau_\pm$.

The quantity ${\mbox{\boldmath$I$}}$ which determines the one-loop
prefactor of our statistical sum (\ref{prefactor0})-(\ref{Y}) reads
in terms of $\varPsi$ as
    \begin{eqnarray}
    &&{\mbox{\boldmath$I$}}=
    2\,\varepsilon_{\cal D}\,
    (\,\varPsi_+\varPsi'_+
    -\varPsi_-\varPsi'_-\,),                 \label{bfI}\\
    &&\varPsi_\pm\equiv
    \varPsi(\tau_\pm,\tau_*),\,\,\,\,
    \varPsi'_\pm\equiv\varPsi'(\tau_\pm,\tau_*). \label{Psis}
    \end{eqnarray}
Because of $g''(\tau_\pm)=0$ the function $\varPsi(\tau,\tau_*)$ is
differentiable at $\tau\to\tau_\pm$, and all the quantities which
enter this expression are well defined. These properties of
$\varPsi(\tau,\tau_*)$ guarantee that (\ref{bfI}) is independent of
an arbitrary choice of the point $\tau_*$, which can be easily
verified by using a simple relation $d\varPsi'_\pm/d\tau_*=-
g'_\pm/g^2(\tau_*)$.

\section{Quadratic part of the action}
The quadratic part of the action can be simplified by a systematic
use of equations of motion for the background (\ref{efeq}) and
    \begin{eqnarray}
    &&\frac1N\frac{\delta\varGamma}{\delta a}
    =\frac{\partial\cal
    L}{\partial a}-\left(\frac{\partial\cal
    L}{\partial a'}\right)'
    -\frac1{a^2}\,\frac{dF}{d\eta}=0,           \label{backgreq2}
    \end{eqnarray}
and their time derivatives (note that (\ref{backgreq2}) is the
derivative of (\ref{efeq}) which is the manifestation of the
time-parametrization invariance). Thus, after integration by parts,
the second order variation in $\delta a$ of the local part of the
action (\ref{action1}) can be transformed by using the
differentiated (and devided by $a'$) version of (\ref{backgreq2}),
    \begin{eqnarray}
    \delta_a^2\varGamma=\oint d\tau\left\{\,{\cal D}\big(\delta a'\big)^2+
    \frac1{a'}\left({\cal D}a''\right)'\delta a^2\right\}+
    \frac{d^2F}{d\eta^2}(\delta_a\eta)^2,
    \end{eqnarray}
where ${\cal D}$ is defined by (\ref{g}), $\delta_a\eta=-\oint
d\tau\,\delta a/a^2$ and the background value of the lapse function
was chosen to be $N=1$.

The functional $\varGamma=\varGamma[\,a,N\,]$ is invariant with
respect to the linearized version of the one-dimensional
diffeomorphism (\ref{diffeo}), $\bar\tau=\tau+f(\tau)$, with the
periodic parameter $f(\tau)$ on a circle
$-\tau_-\leq\tau\leq\tau_+$,
    \begin{eqnarray}
    &&\Delta^f\delta N\equiv
    \overline{\delta N}(\tau)-\delta N(\tau)=-f',   \label{diffeo1}\\
    &&\Delta^f\delta a\equiv
    \overline{\delta a}(\tau)-\delta a(\tau)=-a'f.   \label{diffeo2}
    \end{eqnarray}
Therefore it satisfies the following Ward identity
    \begin{eqnarray}
    \frac{\delta\varGamma}{\delta a}=
    \frac{N}{a'}
    \left(\frac{\delta\varGamma}{\delta N}\right)',
    \end{eqnarray}
which allows one to simplify on shell (\ref{efeq}) the mixed $\delta
N\,\delta a$-variation of the action,
    \begin{eqnarray}
    \delta_N\delta_a\varGamma=\oint d\tau N
    \left(\delta_N\frac{\delta\varGamma}{\delta N}\right)'
    \frac{\delta a}{a'}=-\oint d\tau
    \left(\frac{\delta a}{a'}\right)'({\cal D}a'^2)\,\delta N+
    \frac{d^2F}{d\eta^2}\,\delta_N\eta\,\delta_a\eta,
    \end{eqnarray}
where $\delta_N\eta=\oint d\tau\,\delta N/a$.  Here the integration
by parts in the first term is admissible because the derivative of
the singular at $a'=0$ quantity $\delta a/a'$ is compensated by the
factor $a'^2$.

Similarly
    \begin{eqnarray}
    \delta_N^2\varGamma=\delta_N\oint d\tau \left({\cal L}-\frac{\partial\cal
    L}{\partial a'}\,a'+\frac1a\,\frac{dF}{d\eta}\right)\delta N
    =\oint d\tau \big({\cal D}a'^2\big)\,\delta N^2+
    \frac{d^2F}{d\eta^2}\,(\delta_N\eta)^2,      \label{}
    \end{eqnarray}
and the second order variation of the action in terms of
perturbations $\delta a$ and $\delta N$ finally takes the form
    \begin{eqnarray}
    &&\varGamma_{(2)}=\frac12\,\delta^2\varGamma=
    \frac12\oint d\tau\left\{\,{\cal D}\big(\delta a'\big)^2+
    \frac1{a'}\left({\cal D}a''\right)'\delta a^2\right.\nonumber\\
    &&\qquad\qquad\qquad\qquad
    \left.-2\left(\frac{\delta a}{a'}\right)'({\cal D}a'^2)\,\delta N
    +\big({\cal D}a'^2\big)\,\delta N^2\right\}+
    \frac12\frac{d^2F}{d\eta^2}(\delta\eta)^2,     \label{quadraction0}\\
    &&\delta\eta=\delta_N\eta+\delta_a\eta=\oint
    d\tau\left(\frac{\delta N}a
    -\frac{\delta a}{a^2}\right).              \label{etavariation}
    \end{eqnarray}

Formal integration by parts allows one to convert the integral of
the first three terms to the quadratic form
    \begin{eqnarray}
    \frac12\oint d\tau\,\big({\cal
    D}a'^2\big)\,\varPsi^2,\,\,\,\,\varPsi\equiv\delta
    N-\left(\frac{\delta a}{a'}\right)'
    \end{eqnarray}
in the variable $\varPsi$ which is a local invariant of the
linearized diffeomorphism transformations\footnote{This is a long
wavelength version of the gauge-invariant variable of cosmological
perturbations \cite{Bardeen,Mukhanov}},
(\ref{diffeo1})-(\ref{diffeo2}). This representation is however
illegitimate because the integrand of this form is not integrable in
the vicinity of points with $a'=0$, these divergences being acquired
via divergent total derivative terms. The action simplifies in terms
of another set of variables which were introduced in Sect.2 by
Eqs.(\ref{newvariables1})-(\ref{newvariables2}) with the function
$g=g(\tau)$ given by (\ref{g}).

In terms of these variables the variation of this global degree of
freedom -- the conformal time period of the instanton -- reads as
    \begin{eqnarray}
    \delta\eta=\oint d\tau\, n
    \end{eqnarray}
and, as one can easily check, the local part of the quadratic action
(\ref{quadraction0}) after a number of nonsingular integrations by
parts takes the form
    \begin{eqnarray}
    &&\frac12\oint d\tau\left\{\,{\cal D}\big(\delta a'\big)^2+
    \frac1{a'}\left({\cal D}a''\right)'\delta a^2
    -2\left(\frac{\delta a}{a'}\right)'({\cal D}a'^2)\,\delta N
    +\big({\cal D}a'^2\big)\,\delta N^2\right\}\nonumber\\
    &&\qquad\qquad=
    \frac12\,\varepsilon_{\cal D}\oint
    d\tau\left\{\,\varphi'^2+\frac{g''}g\,\varphi^2
    +\big(\,4ng'+2n'g\big)\varphi+g^2n^2\,\right\}.  \label{part1}
    \end{eqnarray}
Thus we come to the final elegant form of the quadratic action
(\ref{quadraction}), which turns out to be functionally
parameterized by a single function $g$.

\section{The choice of gauge conditions}
\subsection{Admissibility of relativistic gauges}
In terms of the new variables $\phi=(\varphi,n)$ the diffeomorphism
transformations (\ref{diffeo1})-(\ref{diffeo2}) take the form
    \begin{eqnarray}
    \Delta^f n=-{\tilde f}',\,\,\,\,
    \Delta^f\varphi=-g\,\tilde f,\,\,\,\,
    \tilde f\equiv\frac{f}a                \label{diffeo3}
    \end{eqnarray}
with the rescaled parameter of gauge transformations $\tilde f$. The
gauge condition $\chi=\chi(\phi)$ should be such that any field
$\phi$ can be transformed to the representative of the gauge orbit
(denoted by the bar, $\bar\phi$) which satisfies this gauge. In
other words the equation $0=\bar\chi\equiv\chi+\Delta^f \chi$ should
always have a unique solution $f$ for any $\phi$, which means that
the Faddeev-Popov operator $Q=Q(d/d\tau)$ in $\Delta^f\chi\equiv Qf$
should be invertible.

Though locally the equation for $f$ might be solvable, there can be
global obstructions to the existence of such a solution. For
example, take the gauge on the Lagrangian multiplier not containing
its time derivative, $\chi=n$. The transition to this gauge is
achieved via the gauge parameter solving the equation ${\tilde
f}'=n$, which is however not periodic because of an obvious jump
$\tilde f(T)-\tilde f(0)=\oint d\tau n$. On the contrary, the {\em
relativistic} gauge containing a time derivative of $n$, like
$\chi=n'$, has a larger freedom in boundary conditions admitting
periodic solutions for diffeomorphisms. Indeed, the solution of
${\tilde f}''=n'$,
    \begin{eqnarray}
    \tilde f(\tau)=\int_0^\tau d\tau_1\,n_1+C\tau+\tilde f(0),
    \end{eqnarray}
is periodic under the following choice of the integration constant
$C=-\oint d\tau_1\,n_1/T$.

\subsection{Special gauge: positive definiteness of the Euclidean
action} The gauge fixing procedure generally can affect the form of
the action and, in particular, change the convexity property of its
quadratic part in the sector of gauge degrees of freedom. The action
(\ref{quadraction}) prior to gauge fixing is not positive/negative
definite, because the potential term $g''/g$ in the quadratic form
in $\varphi$ is basically negative\footnote{Remember that
$g(\tau)\sim a'(\tau)$ is an oscillating function, and its convexity
in the average is opposite to the sign of its amplitude}, and the
operator ${\mbox{\boldmath$F$}}$ is indefinite. This can be improved
by using the gauge of the form
    \begin{eqnarray}
    \chi=\frac{g''}g\,\varphi+2ng'+n'g=0.    \label{positgauge}
    \end{eqnarray}
Such a gauge is admissible because $\Delta^f\chi=(gf/a)''$, and its
Faddeev-Popov operator
    \begin{eqnarray}
    Q(d/d\tau)f=\frac{d^2}{d\tau^2}\frac{gf}a
    \end{eqnarray}
is invertible under the condition of regularity. Indeed, the only
periodic zero mode of this operator $f\propto a/g$ is singular at
$\tau_\pm$ and should be discarded. Therefore, $Q$ has a regular
periodic Green's function on a circle. With this gauge imposed the
quadratic action (\ref{quadraction}) takes the form with the
reversed sign of $(g''/g)\varphi^2$, whereas all the other terms
become definite in sign (positive or negative depending on
$\varepsilon_{\cal D}=\pm1$)
    \begin{eqnarray}
    &&\varGamma_{(2)}=
    \frac12\,\varepsilon_{\cal D}\oint
    d\tau\,\left\{\varphi'^2-\frac{g''}g\,\varphi^2
    +g^2n^2\right\}+
    \frac12\,\frac{d^2F}{d\eta^2}
    \left(\oint d\tau\, n\right)^2.            \label{positaction}
    \end{eqnarray}
In our concrete model of the CFT driven cosmology we have
    \begin{eqnarray}
    \varepsilon_{\cal D}=-1,\,\,\,\,\frac{d^2F}{d\eta^2}<0,\,\,\,\,
    \frac{g''}g<0,
    \end{eqnarray}
and the action becomes negative definite for real $n$ and $\varphi$.
With $\varphi$ expressed via the gauge (\ref{positgauge}) as
    \begin{eqnarray}
    \varphi=-\frac{(ng^2)'}{g''},   \label{varphiofn}
    \end{eqnarray}
this provides a well defined procedure of Gaussian integration over
$n$ along the imaginary axis contour $n\in [\,-i\infty,+i\infty\,]$.
This is dictated by the original definition of the microcanonical
path integral in the physical spacetime with the Lorentzian
signature \cite{why,tunnel}. Indeed, the Euclidean path integral
(\ref{1}) is the transformed version of the path integral over
Lorentzian signature metrics for the microcanonical statistical sum
in cosmology \cite{why,tunnel}. This definition implies the
integration over imaginary values of the Euclidean gravity lapse
function and signifies an imaginary $n$ -- the alternative
((3+1)-noncovariant) version of the so-called conformal rotation in
Euclidean quantum gravity designed to render its Einstein action
positive-definite \cite{confrotation}.

Unfortunately, the action (\ref{positaction}) with (\ref{varphiofn})
has fourth-order derivatives and is hard to handle. Therefore we
consider below another relativistic gauge. It does not improve the
convexity properties of the action (and thus makes the choice of the
path integration contour trickier), but renders the problem
tractable from the calculational point of view.

\subsection{The relativistic gauge and residual gauge transformations}
In what follows we will use the relativistic gauge
    \begin{eqnarray}
    \chi\equiv n'=0    \label{localgauge}
    \end{eqnarray}
in which the variable $n$ becomes constant, $n=n_0={\rm const}$, and
the action simplifies to
    \begin{eqnarray}
    &&\varGamma_{(2)}=
    \frac12\,\varepsilon_{\cal D}\oint
    d\tau\,\left\{\varphi'^2+\frac{g''}g\,\varphi^2
    +4\,n_0 g'\varphi\right\}\nonumber\\
    &&\qquad\qquad\qquad\qquad\quad
    +\frac12\,\varepsilon_{\cal D}\,n_0^2\,\oint
    d\tau\,g^2+
    \frac12\,\frac{d^2F}{d\eta^2}\,n_0^2 T^2,   \label{finalquadr}
    \end{eqnarray}
where $T$ denotes the full period of the instanton
    \begin{eqnarray}
    T=\oint d\tau.    \label{T}
    \end{eqnarray}

This gauge does not fix the gauge freedom completely, because it
remains invariant under the residual gauge transformations
(\ref{diffeo3}) with a special parameter
$f(\tau)=-a(\tau)\varepsilon$, $\varepsilon={\rm const}$. It has a
simple geometric interpretation -- conformal Killing symmetry of a
generic FRW background. Indeed, with this choice of $f$ the
diffeomorphism of $\delta a$ and $\delta N$
(\ref{diffeo1})-(\ref{diffeo2}) coincides with the local conformal
transformation of these perturbations on the background of
$(a(\tau),N=1)$
    \begin{eqnarray}
    \Delta^\omega\delta a(\tau)
    =\omega(\tau)\,a(\tau),\,\,\,\,
    \Delta^\omega\delta N(\tau)=\omega(\tau)\,N=
    \omega(\tau)                        \label{conformaltransf}
    \end{eqnarray}
with the conformal factor parameter
$\omega(\tau)=a'(\tau)\,\varepsilon$.\footnote{With another choice
of time in the FRW metric, $N\neq 1$, these transformations are
modified by factors of $N$, and the conformal Killing diffeomorphism
is given by $f=-\varepsilon\,a/N$.} This explains the origin of
extra symmetry not fixed by the coordinate gauge $n'=0$ --
conformally non-invariant Einstein and anomaly parts of the action
(\ref{action1}) become invariant under those conformal
transformations which coincide with diffeomorphisms, and this occurs
for the conformal Killing transformation which exists for any FRW
metric.

The parameter $f(\tau)=-a(\tau)\varepsilon$ is regular and periodic,
and therefore this residual gauge freedom cannot be ruled out by
boundary conditions, but rather has to be fixed by an extra gauge
condition. As we will see now this procedure leads to the problem of
quantization of the theory with linearly dependent generators.

In order to formulate the full set of resulting gauges we first
collect all variables and the parameters of transformations --
original local diffeomorphisms $\Delta^f\phi^i$ and residual global
diffeomorphisms
$\Delta^\varepsilon\phi^i\equiv\Delta^f\phi^i\;\big|_{\,f=-a\varepsilon}$
-- into their multiplets
    \begin{eqnarray}
    \phi^i=
    \left[\begin{array}{c}
        \varphi(\tau)\\
        \\
        \;n(\tau)\;
    \end{array}\right],\,\,\,\,
        f^\alpha=
    \left[\begin{array}{c}
        \;\,f(\tau)\;\\
        \\
        \;\varepsilon\;
    \end{array}\right].
    \end{eqnarray}
Here the condensed indices $i$ and $\alpha$ include the continuous
entry $\tau$ along with the discrete entry of the above
2-dimensional columns. In these notations the full set of
transformations reads as $\Delta^f\phi^i=R^i_\alpha f^\alpha$, where
the generators $R^i_\alpha$ form the functional matrix
    \begin{eqnarray}
    R^{i_1}_{\alpha_2}=
    \left[\begin{array}{cc}
        -{\displaystyle\frac{g_1\delta(\tau_1,\tau_2)}{a_2}}
        \;&\;\;\;g_1\;\\
        \\
        \;-{\displaystyle\frac{d}{d\tau_1}
        \frac{\delta(\tau_1,\tau_2)}{a_2}}
        \;\;&0
    \end{array}\right],               \label{matrixgenerator}
    \end{eqnarray}
in which numerical labels of condensed indices imply different time
coordinates $\tau_1$ and $\tau_2$ of their continuous entries. The
contraction over $\alpha$ in $\Delta^f\phi^i=R^i_\alpha f^\alpha$
implies the integration over continuous time entries of $\alpha$
along with the summation over its discrete entries. The generators
of both transformations $\Delta^f\phi^{i_1}=\oint
d\tau_2\,R^{i_1}_{f_2}\,f_2$ and
$\Delta^\varepsilon\phi^{i_1}=R^{i_1}_\varepsilon\,\varepsilon$
(with $R^{i_1}_{f_2}$ and $R^{i_1}_\varepsilon$ given by the first
and the second column of the matrix (\ref{matrixgenerator}))
obviously form a linearly dependent set satisfying
    \begin{eqnarray}
    \oint d\tau_2\,R^{i_1}_{f_2} a(\tau_2)
    +R^{i_1}_\varepsilon=0.        \label{linearrelation}
    \end{eqnarray}
In condensed notations this relation reads as
      \begin{eqnarray}
        R^i_\alpha Z^\alpha=0.       \label{generatordependence}
    \end{eqnarray}
with the coefficients
   \begin{eqnarray}
        Z^\alpha=
    \left[\begin{array}{c}
        \;\,a(\tau)\;\\
        \\
        \;1\;
    \end{array}\right].
    \end{eqnarray}

As an additional gauge condition fixing the residual gauge
transformation one can take a (functional) orthogonality of the
quantum field $\varphi$ to the $\varepsilon$-transformation of
$\Delta^\varepsilon\varphi$. This is the following global gauge
condition
    \begin{eqnarray}
    \oint d\tau\,g\varphi=0,
    \end{eqnarray}
which of course is admissible because the relevant Faddeev-Popov
operator $Q$, characterizing the transformation of this gauge
condition $\Delta^\varepsilon\oint d\tau\,g\varphi=Q\varepsilon$, is
given by (\ref{FPzeromode}) and is invertible for any identically
nonvanishing function $g$.

This global gauge together with the local gauge conditions
(\ref{localgauge}) forms the following set
    \begin{eqnarray}
    \chi^\alpha=
    \left[\begin{array}{c}
        \;n'(\tau)\\
        \\
        \;\displaystyle\oint d\tau\,g\varphi\;
    \end{array}\right]                           \label{fullgauge}
    \end{eqnarray}
which gives rise to the full block-structure Faddev-Popov operator,
$\Delta^f\chi^\alpha=Q^\alpha_\beta\,f^\beta$,
    \begin{eqnarray}
    Q^\alpha_\beta=
    \left[\begin{array}{cc}
        \;-{\displaystyle\frac{d^2}{d\tau_1^2}
        \frac{\delta(\tau_1,\tau_2)}{a_2}}
        \,\;\;&\;\;0\;\;\\
        \\
        \;-{\displaystyle\frac{g_2^2}{a_2}}\;\;&Q
    \end{array}\right].                          \label{fullFP}
    \end{eqnarray}
This operator,
$Q^\alpha_\beta=(\delta\chi^\alpha/\delta\phi^i)R^i_\beta$, is
degenerate, because of the linear dependence of generators
(\ref{generatordependence}) and has as right zero-eigenvalue
eigenvector the coefficients $Z^\alpha$
    \begin{eqnarray}
    Q^\alpha_\beta Z^\beta=0.
    \end{eqnarray}
Inevitable left zero-eigenvalue eigenvector of this operator follow
from the properties of the full set of gauge conditions
(\ref{fullgauge}). In view of $\oint d\tau\,n'=0$ this set,
similarly to the generators, is redundant and satisfies the
functional linear dependence relation $\bar Z_\alpha\chi^\alpha=0$
with
    \begin{eqnarray}
    \bar Z_\alpha= \left[\begin{array}{cc}
        \;1\;&\;0\;
    \end{array}\right].
    \end{eqnarray}
Here, as above, contraction of condensed indices in the
time-dependent entry of the full index $\alpha=(\tau,\varepsilon)$
includes the $\tau$-integration. These coefficients form the left
zero-eigenvalue eigenvector of (\ref{fullFP})
    \begin{eqnarray}
    \bar Z_\alpha Q^\alpha_\beta=0.
    \end{eqnarray}

A regular technique of handling gauge theories with linear dependent
generators and redundant sets of gauge conditions is known as a part
of the BV formalism \cite{BFV}, and we briefly present it in the
next section.

\section{BV formalism for reducible gauge
theories: one-loop approximation} The BV formalism of \cite{BFV}
suggests a quantization method for a generic gauge theory whose
action $S[\,\phi\,]$ is invariant under gauge transformations with
linearly dependent generators $R^i_\alpha$
    \begin{eqnarray}
    &&\frac{\delta S[\,\phi\,]}{\delta\phi^i}\,R^i_\alpha=0,\\
    &&R^i_\alpha Z^\alpha_a=0.
    \end{eqnarray}
The coefficients of these linear dependence relations $Z^\alpha_a$
are labeled by a condensed index $a$. For simplicity we present here
the case when all gauge fields, parameters of gauge transformations
and these zero vectors $Z^\alpha_a$ are bosonic classically
commuting variables and when these zero vectors are linearly
independent themselves (the first-stage reducibility with the rank
of the matrix $Z^\alpha_a$ coinciding with the range of the index
$a$). The general case of boson-fermion systems of arbitrary stage
of reducibility is fully considered in \cite{BFV}.

The gauge invariance of the action implies imposing the set of
gauges
    \begin{eqnarray}
    \chi^\alpha=\chi^\alpha(\phi),
    \,\,\,\,\,\bar Z_\alpha^a\chi^\alpha=0 \label{redundantgauge}
    \end{eqnarray}
which should also be redundant not to overconstrain the system,
which is actually invariant under the set of transformations whoes
number is less than the range of index $\alpha$ -- the rank of the
matrix $R^i_\alpha$. Thus the numbers of zero vectors of generators
and zero vectors of constraints $\bar Z_\alpha^a$ (the range of the
index $a$) should coincide. The usual Faddeev-Popov operator for
this set of gauges
    \begin{eqnarray}
    Q^\alpha_\beta=\frac{\delta\chi^\alpha}{\delta\phi^i}
    R^i_\beta
    \end{eqnarray}
is itself degenerate and has as right and left zero vectors
$Z^\beta_b$ and $\bar Z_\alpha^a$ respectively. This means that the
Faddeev-Popov ghosts $(C^\beta,\bar C_\alpha)$ which should generate
${\rm Det}\,Q^\alpha_\beta$ in the gauge-fixed path integral also
become gauge fields and require auxiliary gauge fixing. Moreover,
the redundancy of gauge conditions implies that their formal delta
function
$\delta[\,\chi\,]=\prod_\alpha\delta[\,\chi^\alpha\,]\sim\delta[\,0\,]$
is not well-defined and requires regularization.

For the first-stage reducible theories this quantum gauge-fixing
procedure is achieved by a special extension of the set of ghost
fields and Lagrange multipliers \cite{BFV}
    \begin{eqnarray}
    \varPhi_{\rm g}=(C^\alpha;
    \bar C_\alpha,\pi_\alpha)\to\varPhi_{\rm
    g}=(C^\alpha,C^a;\bar C_\alpha,\pi_\alpha,
    \bar C_a,\pi_a,E^a,P^a)           \label{ghostsector}
    \end{eqnarray}
in the effective (gauge-fixed) action $S_{\rm
eff}[\phi^i,\varPhi_{\rm g}]$ and the corresponding path integral
    \begin{eqnarray}
    &&Z=\int D\phi\,D\varPhi_{\rm g}\,\exp\Big(-S_{\rm
    eff}\Big),\\
    &&S_{\rm eff}[\phi^i,\varPhi_{\rm g}]=S+\bar
    C_\alpha Q^\alpha_\beta C^\beta+\bar C_a \big(\omega^a_\alpha
    Z^\alpha_b\big)\,C^b\nonumber\\
    &&\qquad\qquad
    +(\chi^\alpha+\sigma^\alpha_a E^a)\pi_\alpha
    +\pi_a\omega^a_\alpha C^\alpha+\bar C_\alpha
    \sigma^\alpha_a P^a.
    \end{eqnarray}
Here $(\bar C_\alpha,C^\beta,\pi_a,P^a)$ are Grassman anti-commuting
variables whereas the rest are bosonic ones. The meaning of
additional Lagrange multipliers $\pi_a$ and $P^a$ is that they
generate delta functions of gauge conditions $\omega^a_\alpha
C^\alpha$ and $\bar C_\alpha\sigma^\alpha_a$ for the original
Faddeev-Popov ghosts $(C^\alpha,\bar C_\alpha)$ which now are gauge
fields themselves. $\omega^a_\alpha$ and $\sigma^\alpha_a$ are
arbitrary parameters of these gauges. The variables $C^b$ and $\bar
C_a$ are the corresponding ghosts for the ghost $C^\alpha$. Finally,
the integration over the Lagrange multiplier $\pi_\alpha$ of the
original Faddeev-Popov scheme generates the delta function of the
gauges $\chi^\alpha+\sigma^\alpha_a E^a$ which are modified by the
contribution of the so-called extraghost $E^a$ \cite{BFV}. These
modified gauges are no longer linearly dependent, which makes their
delta function well-defined.

Thus, the integration over the ghost sector gives
    \begin{eqnarray}
    &&\int D\varPhi_{\rm g}\,\exp\Big(-S_{\rm
    eff}\Big)\nonumber\\
    &&\qquad\qquad\qquad\quad
    =\exp\Big(-S\Big)\,{\rm Det}
    \left[\begin{array}{cc}
        \;\;Q^\alpha_\beta\;\;&\;\sigma^\alpha_b\;\\
        \\
        \;\;\omega^a_\beta\;\;&0
    \end{array}\right]\,\frac1{{\rm Det}\big(\omega^a_\alpha
    Z^\alpha_b\big)}\,\int DE^a\,
    \delta\big[\,\sigma^\alpha_a E^a+\chi^\alpha\big]\nonumber\\
    &&\\
    &&\qquad\qquad\qquad\quad
    =\exp\Big(-S\Big)\,
    \frac{{\rm Det}{\cal F}^\alpha_\beta}
    {{\rm Det}\,q^a_b\,
    {\rm Det}\,\overline q^a_b}\,
    \int DE^a\,
    \delta\big[\,\sigma^\alpha_a E^a
    +\chi^\alpha\big]\,
    \,\big(\,{\rm Det}\,
    \overline q^a_b\big),          \label{intoverg}
    \end{eqnarray}
where now the ghost operator ${\cal F}^\alpha_\beta$ is a
gauge-fixed version of the degenerate $Q^\alpha_\beta$,
    \begin{eqnarray}
    {\cal F}^\alpha_\beta=Q^\alpha_\beta
    +\sigma^\alpha_a\omega^a_\beta.      \label{gfixedFP}
    \end{eqnarray}
The determinants of matrices
    \begin{eqnarray}
    &&q^a_b=\omega^a_\alpha \,Z^\alpha_b,     \label{q0}\\
    &&\bar q^a_b=
    \bar Z_\alpha^a\,\sigma^\alpha_b        \label{barq0}
    \end{eqnarray}
represent the Faddeev-Popov ghost factors for the original ghost
fields (note that their double nature corresponds to two different
zero modes of $Q^\alpha_\beta$ -- its right and left zero vectors).
A simple exercise using these zero modes shows that the ratio of
determinants in (\ref{intoverg}) is independent of the choice of
gauge parameters $\sigma^\alpha_b$ and $\omega^a_\alpha$ (see
Appendix A)
    \begin{eqnarray}
    \delta_{(\omega,\sigma)}
    \frac{{\rm Det}{\cal F}^\alpha_\beta}
    {{\rm Det}\,q^a_b\,
    {\rm Det}\,\overline q^a_b}=0.  \label{omegavariation}
    \end{eqnarray}
Also in the case of a redundant set of gauge conditions
(\ref{redundantgauge})
    \begin{eqnarray}
    \delta_{\sigma}\Big\{\,
    \delta\big[\,\sigma^\alpha_a E^a+\chi^\alpha\big]\,
    \,\big(\,{\rm Det}\,
    \overline q^a_b\big)\Big\}=0.            \label{sigmavariation}
    \end{eqnarray}
This makes the ghost sector of the path integral independent of the
choice of functions $(\sigma^\alpha_b,\omega^a_\alpha )$ fixing
additional gauge symmetries associated with the linear dependence of
generators.

\section{Gauge and ghost sector}
We now apply the above procedure to our model. First note that the
range of the index $a$ here is just one discrete value corresponding
to one linear dependence relation (\ref{generatordependence}), so
that we will omit this index at al. Regarding additional gauge
parameters $\sigma^\alpha_b=\sigma^\alpha$ and
$\omega^a_\alpha=\omega_\alpha$, there exists a convenient choice
which renders the Faddeev-Popov operator (\ref{gfixedFP}) a
block-diagonal structure. They read as the following 2-dimensional
column and row
   \begin{eqnarray}
        \sigma^\alpha=
    \left[\begin{array}{c}
        \;\,\sigma(\tau)\;\\
        \\
        \;1\;
    \end{array}\right],\;\;\;\;\;
    \omega_\alpha= \left[\begin{array}{cc}
        \;\,\displaystyle\frac{g^2}a\;\,&\;0\;
    \end{array}\right]                            \label{sigmaalpha}
    \end{eqnarray}
and in view of (\ref{fullFP}) yield the operator
    \begin{eqnarray}
    {\cal F}^\alpha_\beta=
    \left[\begin{array}{cc}
        \;{\cal F}(\tau_1,\tau_2)\,\displaystyle\frac1{a_2}
        \,\;\;&\;\;0\;\;\\
        \\
        \;0\;\;&Q
    \end{array}\right],\,\,\,\,\,
    {\cal F}(\tau,\tau')=-\frac{d^2}{d\tau^2}\,\delta(\tau,\tau')
    +\sigma(\tau)\,g^2(\tau').                          \label{fullFP1}
    \end{eqnarray}
The function $\sigma(\tau)$ is rather arbitrary, and should be
chosen to guarantee the invertibility of the corresponding ghosts
for ghosts operators
      \begin{eqnarray}
        &&\bar q \,\equiv\, \bar Z_\alpha\sigma^\alpha
        =\oint d\tau\,\sigma(\tau),     \label{barq}\\
        &&q\,\equiv\,\omega_\alpha Z^\alpha
        =\oint d\tau\,g^2(\tau)=Q.       \label{q}
    \end{eqnarray}
This is easily attained with a generic positive-definite
$\sigma(\tau)$. Therefore, the calculation of the determinant of
(\ref{fullFP1}) reduces to that of ${\rm Det}\,{\cal F}(\tau,\tau')$
which can be recovered from the variational equation
    \begin{eqnarray}
    \delta_{\sigma}\ln\Big[\;{\rm Det}\,
    {\cal F}(\tau,\tau')\,\Big]=\oint d\tau_1\,\delta\sigma_1\oint
    d\tau_2\,g_2^2\,G(\tau_2,\tau_1).
    \end{eqnarray}
Here $G(\tau_2,\tau_1)=\delta f(\tau_2)/\delta J(\tau_1)$ is the
inverse of the operator ${\cal F}(\tau_1,\tau_2)$ or the Green's
function of the problem
    \begin{eqnarray}
    -f''(\tau)+\sigma(\tau)
    \oint d\tau_1\,g^2(\tau_1)\,f(\tau_1)=J(\tau)  \label{problem}
    \end{eqnarray}
for the function $f(\tau)$ periodic on a circle (with periodic
derivatives). In view of the periodicity of $f'$ the integration of
this equation gives $\oint d\tau\,g^2 f=\oint d\tau\,J(\tau)/\bar
q$, so that the variation of the functional determinant above reads
    \begin{eqnarray}
    &&\delta_{\sigma}\ln\Big[\;{\rm Det}\,
    {\cal F}(\tau,\tau')\,\Big]=\oint
    d\tau\,g^2 f\,\Big|_{\;J=\delta\sigma}\nonumber\\
    &&\qquad\qquad\qquad\qquad\quad=
    \frac1{\overline q}\oint d\tau\,
    \delta\sigma(\tau)=\delta_\sigma\ln\overline q.
    \end{eqnarray}

The functional dependence of this determinant on $g(\tau)$ follows
from a similar variation
    \begin{eqnarray}
    &&\delta_{g}\ln\Big[\;{\rm Det}\,
    {\cal F}(\tau,\tau')\,\Big]=\oint d\tau_1\,\sigma_1\oint
    d\tau_2\,\delta(g_2^2)\,G(\tau_2,\tau_1)\nonumber\\
    &&\qquad\qquad\qquad\qquad\quad=\oint
    d\tau\,\delta(g^2)\, f\,\Big|_{J=\sigma}=\delta_g\ln Q.
    \end{eqnarray}
Here we have used the fact that the solution of the problem
(\ref{problem}) for a special choice of the source
$J(\tau)=\sigma(\tau)$ reads as $f\,|_{J=\sigma}=1/Q$. Indeed, with
this choice $\oint d\tau\,g^2f=1$, and the resulting equation
$f''=0$ has as a periodic solution the constant -- the inverse of
$\oint d\tau\, g^2=Q$. Therefore
    \begin{eqnarray}
    &&{\rm Det}\,
    {\cal F}(\tau,\tau')={\rm const}\times\bar q\,Q,
    \end{eqnarray}
so that the ratio of determinants ${\rm Det}{\cal F}^\alpha_\beta/
{\rm Det}\,q^a_b\,{\rm Det}\,\overline q^a_b$ in (\ref{intoverg})
which can be regarded as the gauge-independent definition of the
restricted functional determinant ${\rm Det}_*{\mbox{\boldmath$Q$}}$
of (\ref{statsum1}) reads as
    \begin{eqnarray}
    {\rm Det}_*{\mbox{\boldmath$Q$}}
    \equiv\frac{{\rm Det}{\cal F}^\alpha_\beta}
    {q\,\overline q}={\rm const}\times
    Q\,\Big(\prod\limits_\tau a(\tau)\Big)^{-1}. \label{detFP}
    \end{eqnarray}

The delta function of gauge conditions (\ref{fullgauge}) modified by
the extra ghost $E^a=E$ with the choice of gauge parameters
(\ref{sigmaalpha}) reads as
    \begin{eqnarray}
    \delta\big[\,\sigma^\alpha E+\chi^\alpha\big]=
    \delta\!\left(E+\oint d\tau\, g\varphi\right)\,
    \delta\big[\,n'(\tau)+\sigma(\tau)E\,\big].    \label{deltagauge}
    \end{eqnarray}
The support of the second delta function here is given by the
solution of the equation $n'(\tau)+\sigma(\tau)E=0$, which implies
in view of the periodicity of $n(\tau)$ and non-degeneracy of $\bar
q$, that separately $E=0$ and $n'=0$. Therefore the functional
integral over $n(\tau)$ of this delta function with any functional
$\varPhi[\,n(\tau)\,]$ should reduce to the ordinary integral over
the constant mode of $n(\tau)$, $n_0={\rm const}$, and be
proportional to $\delta(\,E\,)$. As shown in the Appendix B this is
indeed the case
    \begin{eqnarray}
    &&\int Dn\,dE\;
    \delta\big[\,n'(\tau)+\sigma(\tau)E\,\big]\,
    \varPhi\big[\,n(\tau),E\,\big]\,\overline q=
    {\rm const}\times T
    \int\limits_{-\infty}^{+\infty}
    dn_0\,\varPhi[\,n_0,0\,].                 \label{integralovernE}
    \end{eqnarray}
This, in particular, confirms the relation (\ref{sigmavariation}).

\section{Metric variables sector}
We apply now the path integral (\ref{intoverg}) to our model with
$S[\,\phi\,]=\varGamma_{(2)}[\,\varphi,n\,]$. Using the expressions
for the gauge-fixed ghost contribution (\ref{detFP}) and the delta
function of the full set of gauge conditions (\ref{deltagauge}) in
this integral we have on account of (\ref{integralovernE})
    \begin{eqnarray}
    &&P=\int D\phi\,D\varPhi_{\rm g}\,\exp\Big(-S_{\rm
    eff}\Big)\nonumber\\
    &&\nonumber\\
    &&\qquad\quad
    =\int D\varphi\,Dn\,
    \exp\Big(-\varGamma_{(2)}[\,\varphi,n\,]\Big)\,
    \frac{{\rm Det}{\cal F}^\alpha_\beta}{q}\,
    \int dE\,\delta\big[\,\sigma^\alpha
    E+\chi^\alpha\big]\nonumber\\
    &&\nonumber\\
    &&\qquad\quad
    ={\rm const}\times \,T\int_{-\infty}^\infty dn_0\int D\varphi\,
    \exp\Big(-\varGamma_{(2)}[\,\varphi,n_0\,]\Big)\,
    \delta\!\left(\oint d\tau\, g\varphi\right)\,Q.
    \end{eqnarray}
Here the local factor in the canonical integration measure
\cite{BarvU}
    \begin{eqnarray}
    D\phi=D(\delta a)\,D(\delta N)\prod_\tau \big|\,{\cal
    D}(\tau)\,\big|^{1/2}=D\varphi\,Dn\,\prod_\tau a(\tau)
    \end{eqnarray}
gets canceled by the local factor of (\ref{detFP}), and also all
factors of $\overline q$ cancel out, as they should due to the gauge
independent nature of the gauge-fixing procedure.

With the quadratic action (\ref{finalquadr}) essentially simplified
for a constant $n(\tau)=n_0$, this integral takes the form
    \begin{eqnarray}
    P=T\int_{-\infty}^\infty dn_0\,
    \exp\left\{\frac12\,n_0^2\,
    \left(-\varepsilon_{\cal D}\,Q
    -\frac{d^2F}{d\eta^2}\,
    T^2\right)\right\}\,K(n_0)   \label{nintegral}
    \end{eqnarray}
where $Q$ is just the Faddeev-Popov factor (\ref{FPzeromode}), $T$
is a full period (\ref{T}) of the Euclidean time and $K(n_0)$ is the
following path integral
    \begin{eqnarray}
    &&K(n_0)=\int D\varphi\;
    \delta\Big(\oint d\tau\, g\varphi\Big)\,Q\,
    \exp\left\{-\varepsilon_{\cal D}\oint
    d\tau\,\Big(\,\frac12\,\varphi\,{\mbox{\boldmath$F$}}\varphi
    +2\,n_0\, g'\varphi\Big)\,\right\}.         \label{I}
    \end{eqnarray}
Here ${\mbox{\boldmath$F$}}$ is the operator (\ref{operator}) -- the
kernel of the quadratic in $\varphi$ part of
$\varGamma_{(2)}[\varphi,n]$. It has as a zero mode the function
$g(\tau)$ -- the generator of the residual transformation
$\Delta^\varepsilon\varphi=\varepsilon g$. In fact $K(n_0)$ is the
Faddeev-Popov path integral with this residual transformation gauged
out by the auxiliary gauge\footnote{Note that even the presence of
the source term for $\varphi$ in the action does not break its
invariance under the $\varepsilon$-transformation, because for a
constant $n_0$ it transforms by a total derivative term $\sim gg'$.}
-- the lower entry of (\ref{fullgauge}). Note that, modulo the
source term linear in $\varphi$, this path integral is just the
definition of the functional determinant of the degenerate operator
${\mbox{\boldmath$F$}}$ on the subspace of its non-zero eigenmodes,
introduced in (\ref{1}), $({\rm
Det}_*{\mbox{\boldmath$F$}})^{-1/2}$.

Representing the delta function of this gauge via the integral over
the Lagrangian multiplier $\pi$ we get the Gaussian path integral
over $\varPhi=(\varphi(\tau),\pi)$ with the new effective action
    \begin{eqnarray}
    &&K(n_0)=Q\int D\varPhi\,\exp\Big(-\varepsilon_{\cal D}\,
    S_{\rm eff}[\,\varPhi; J\,]\,\Big)\,
    \Big|_{\; J=-2n_0 g'}\nonumber\\
    &&\qquad\qquad\qquad=Q\,\Big({\rm Det}\,
    \mathbb{F}\Big)^{-1/2}
    \exp\Big(-\varepsilon_{\cal D}\,
    S_{\rm eff}[\,\varPhi; J\,]\,\Big)
    \Big|_{\;\rm on\, shell},                         \label{I2}\\
    &&S_{\rm eff}[\,\varPhi;J\,]=
    \oint
    d\tau\,\Big(\,\frac12\,\varphi
    {\mbox{\boldmath$F$}}\varphi-i\pi g\varphi
    -J\,\varphi\Big).
    \end{eqnarray}
Here $\mathbb{F}$ is the Hessian of this action (up to a sign factor
$\varepsilon_{\cal D}=\pm1$) with respect to the function
$\varphi(\tau)$ and the numerical variable $\pi$
    \begin{eqnarray}
    \mathbb{F}=
    \frac{\delta^2S_{\rm eff}}
    {\delta\varPhi_1 \delta\varPhi_2}=
    \left[\,\begin{array}{cc} \;{\mbox{\boldmath$F$}}\,
    \delta(\tau_1,\tau_2)&\,\,\, -i g(\tau_1)\,\\
    &\\
    -i g(\tau_2)&0\end{array}\,\right]     \label{matrixF}
    \end{eqnarray}
(note the position of time entries associated with the variables
$\varPhi_1=(\varphi(\tau_1),\pi)$ and
$\varPhi_2=(\varphi(\tau_2),\pi)$), and the onshell condition here
implies the evaluation of the action at its stationary configuration
-- the periodic solution of the following variational problem with
the source $J=-2\,n_0 g'$,
    \begin{eqnarray}
    &&{\mbox{\boldmath$F$}}
    \varphi(\tau)-i\pi g(\tau)-J(\tau)=0,    \label{first}\\
    &&i\oint d\tau\,g\varphi=0.                 \label{second}
    \end{eqnarray}

\subsection{The solution for the metric perturbation and its action}
Multiplying Eq.(\ref{first}) by $g(\tau)$ and integrating by parts
one finds on account of $\mbox{\boldmath$F$}g=0$ the value of $\pi$,
$\pi=-i\oint d\tau\,gJ/Q$, and a new equation for $\varphi$ with a
modified source $\tilde J(\tau)$
    \begin{eqnarray}
    &&{\mbox{\boldmath$F$}}
    \varphi(\tau)=\tilde J(\tau),         \label{first0}\\
    &&\tilde J(\tau)\equiv J(\tau)-
    \frac{g(\tau)}Q\oint d\tau_1\,g(\tau_1)J(\tau_1)
    ,\,\,\,\,
    \oint d\tau\,g\tilde J\equiv 0.         \label{tildeJ}
    \end{eqnarray}
This source $\tilde J(\tau)$ is functionally orthogonal to the zero
mode $g(\tau)$ -- the property that guarantees the existence of the
solution of this equation whose left hand side is also orthogonal to
$g$. Thus, the problem reduces to the solution of
    \begin{eqnarray}
    &&{\mbox{\boldmath$F$}}
    \tilde\varphi(\tau)=\tilde J(\tau),    \label{equation}\\
    &&\oint d\tau\, g\tilde\varphi=0,          \label{gauge}
    \end{eqnarray}
in terms of which the on shell exponential in (\ref{I2}) reads
    \begin{eqnarray}
    &&S_{\rm eff}[\,\varphi(\tau),\pi; J\,]\,
    \Big|_{\;\rm on\, shell}=
    n_0\oint
    d\tau\,\tilde\varphi\,g'\,
    \Big|_{\;\displaystyle J=-2n_0 g'}.    \label{onshellexponential}
    \end{eqnarray}
Below we present the solution $\tilde\varphi$ along with the Green's
function of the problem (\ref{equation})-(\ref{gauge}), which give
the answer for the exponential and preexponential factor of
$K(n_0)$.

For this we will need an explicit parametrization of the
$\tau$-range associated with the oscillatory nature of the functions
$g(\tau)$ and $a(\tau)$. It was introduced in Sect.2 in the form of
a circle of the circumference $T=2(\tau_+-\tau_-)$, $\tau_-\equiv
0$, parameterized by $\tau$ in the range $-\tau_+\leq\tau\leq\tau_+$
with the points $\pm\tau_+$ identified. The antipodal points on the
circle $\tau_\pm$ are distinguished by the fact that they represent
two first degree zeros of $g(\tau)$ and correspond to the maximum
and minimum of the scale factor.

Then the solution of (\ref{equation})-(\ref{gauge}) can be looked
for as a linear combination of the partial solution of the
inhomogeneous equation (\ref{equation})
    \begin{eqnarray}
    &&\varPhi(\tau)=-g(\tau)\int_0^\tau
    \frac{dy}{g^2(y)}\int_0^{y} d\tau'\,g\tilde
    J(\tau')=-\int_0^\tau d\tau'\,\varPsi(\tau,\tau')\,
    g(\tau')\tilde J(\tau').               \label{Phi}
    \end{eqnarray}
and the two basis functions of ${\mbox{\boldmath$F$}}$ --- the
periodic function $g(\tau)$ and the non-periodic
$\varPsi(\tau,\tau_*)$ with some $\tau'=\tau_*>0$, defined by
(\ref{Psi}). For a negative $\tau$ the role of this second basis
function will be played by $\varPsi(-\tau,\tau_*)$ which also
satisfies the equation
${\mbox{\boldmath$F$}}\varPsi(-\tau,\tau_*)=0$ in view of the odd
nature of $g(\tau)$. The role of these basis functions is to fix the
lack of periodicity of the partial solution $\varPhi(\tau)$, which
itself is continuous on a circle, $\varPhi(-\tau_+)=\varPhi(\tau_+)$
(this is guaranteed by the orthogonality of the source to $g$,
$\oint d\tau\,g\tilde J=0$), but its derivative is discontinuous at
$\tau=\pm(\tau_+\!-0)$, $\varPhi'(\tau_+)-\varPhi'(-\tau_+)\neq 0$.
As shown in the accompanying paper \cite{det} the solution of
(\ref{equation})-(\ref{gauge}) reads
    \begin{eqnarray}
    &&\tilde\varphi(\tau)=\varPhi(\tau)
    +C\,\varPsi(\,|\,\tau\,|,\tau_*)
    +D_+\,g(\tau)\,\theta(\tau)
    +D_-\,g(\tau)\,\theta(-\tau),            \label{ansatz}
    \end{eqnarray}
where
    \begin{eqnarray}
    &&C=-\frac12\,
    \frac{\varPsi_+}{\varPsi_+\varPsi'_+
    -\varPsi_-\varPsi'_-}\,
    \Big(\,\varPhi'(\tau_+)
    -\varPhi'(-\tau_+)\Big),          \label{Cpm}\\
    &&D_\pm =-\frac1Q\oint d\tau\,g(\tau)\varPhi(\tau)
    \mp\frac12\,\varPsi_+\,
    \frac{\varPsi_-\,\varPsi'_-}
    {\varPsi_+\varPsi'_+
    -\varPsi_-\varPsi'_-}\,
    \Big(\,\varPhi'(\tau_+)
    -\varPhi'(-\tau_+)\Big)         \label{D-D}
    \end{eqnarray}
Here $\varPsi_\pm$ and $\varPsi'_\pm$ are given by (\ref{Psis}), and
in the denominators of the above expressions one easily recognizes
the basic ingredient (\ref{bfI}), ${\mbox{\boldmath$I$}}$, of our
final result for the one-loop statistical sum.

The knowledge of $\tilde\varphi(\tau)$ allows one to find the
exponential of (\ref{onshellexponential}). With the source
$J=-2n_0g'$ we have
    \begin{eqnarray}
    \varPhi(\tau)=n_0 g(\tau)\tau,\,\,\,\,\,C=
    \varepsilon_{\cal D}\,n_0\,\frac{T}{\mbox{\boldmath$I$}},
    \end{eqnarray}
whereas the $D$-terms of (\ref{ansatz}) do not contribute to this
quantity. What remains finally readsin view of the symmetry
$g'(\tau)=g'(-\tau)$ as
    \begin{eqnarray}
    &&S_{\rm eff}[\,\varphi(\tau),\pi; J\,]\,
    \Big|_{\;\rm on\, shell}=
    n_0\oint d\tau\,g'\,\varPhi
    +2n_0\,C \int_0^{\tau_+}
    d\tau\,g'\,\varPsi(\tau,\tau_*)\nonumber\\
    &&\qquad\qquad\qquad\qquad\qquad\quad
    =-\frac12\,n_0^2
    \left(\,Q+\varepsilon_{\cal D}\,
    \frac{T^2}{\mbox{\boldmath$I$}}\right). \label{onshellexponential1}
    \end{eqnarray}

The degeneration of ${\mbox{\boldmath$I$}}$ to zero obviously leads
to singularity of the coefficients $C$ and $D_\pm$ above and
indicates the presence of an additional zero mode of the operator
${\mbox{\boldmath$F$}}$. As shown in \cite{det} for
$\mbox{\boldmath$I$}=0$, indeed, the function
$\mbox{\boldmath$\varPsi$}(\tau)=\varPsi(\,|\,\tau\,|,\tau_*)
-2\varPsi_-\varPsi'_-\theta(-\tau)\,g(\tau)$ turns out to be the
second zero mode of ${\mbox{\boldmath$F$}}$. However, this mode does
not generate any additional residual symmetry of the action because
the source term for $\varphi$ in the action of (\ref{I}) is not
invariant under the shift of $\varphi(\tau)$ by
$\mbox{\boldmath$\varPsi$}(\tau)$, $\oint
d\tau\,g'(\tau)\,\mbox{\boldmath$\varPsi$}(\tau)=-T/2\neq 0$.
Therefore, no additional gauge fixing is needed, and the prefactor
will stay well defined also in the limit of
${\mbox{\boldmath$I$}}\to 0$, as we will shortly see below.

\subsection{One-loop prefactor from metric perturbations}
The derivation of the prefactor is done in much detail in the
accompanying paper \cite{det}. Here we only present main steps of
this derivation. For this we need the Green's function $\mathbb{G}$
of the matrix valued operator (\ref{matrixF}),
$\mathbb{F}\mathbb{G}=\mathbb{I}$. It reads
    \begin{eqnarray}
    &&\mathbb{G}=
    \left[\,\,\begin{array}{cc}G(\tau,\tau')\,&\,\,\,
    {\displaystyle \frac{\textstyle i g(\tau)}Q}\,\\
    {\displaystyle \frac{\textstyle i g(\tau')}Q}
    &0\end{array}
    \,\right],                                    \label{matrixG}
    \end{eqnarray}
where the Green's function in the diagonal block satisfies the
system of equations
    \begin{eqnarray}
    &&{\mbox{\boldmath$F$}}
    G(\tau,\tau')=\delta(\tau,\tau')
    -\frac{g(\tau)\,g(\tau')}Q,             \label{noomega}\\
    &&\oint d\tau\,g(\tau)\,G(\tau,\tau')=0,
    \end{eqnarray}
indicating that it is the inverse of the operator
$\mbox{\boldmath$F$}$ on the subspace orthogonal to its zero mode.
One can show \cite{det} that it directly expresses in terms of the
Green's function $\tilde G(\tau,\tau')$ of the problem
(\ref{equation})-(\ref{gauge}),
    \begin{eqnarray}
    &&G(\tau,\tau')=
    \tilde G(\tau,\tau')-
    \oint d\tau_1\,\tilde G(\tau,\tau_1)\,
    \frac{g(\tau_1)\,g(\tau')}Q\,.          \label{G}
    \end{eqnarray}
The latter, in its turn, can be read off (\ref{ansatz}), because
$\varPhi$, $C$ and $D_\pm$ are all linear in $\tilde J$, $\tilde
G(\tau,\tau')=\delta\tilde\varphi(\tau)/\delta\tilde J(\tau')$
\cite{det}.

This Green's function (\ref{matrixG}) allows one to find a one-loop
prefactor of (\ref{I2}) via the variational equation $\delta\ln {\rm
Det}\,\mathbb{F}={\rm
    Tr}\,\big(\delta\mathbb{F}\,
    \mathbb{G}\big)$.
A rather lengthy calculation of this quantity in \cite{det} gives
$\delta\ln {\rm Det}\,\mathbb{F}=
    \delta\ln|\,\mbox{\boldmath$I$}\,|+2\delta\ln Q$
with $\mbox{\boldmath$I$}$ defined by (\ref{bfI}). Therefore the
prefactor of $K(n_0)$, which actually serves as the definition of
the restricted functional determinant of $\mbox{\boldmath$F$}$
introduced in Sect.2, equals
    \begin{eqnarray}
    ({\rm Det_*}{\mbox{\boldmath$F$}}\,)^{-1/2}
    \equiv\Big({\rm Det}\,\mathbb{F}\Big)^{-1/2}
    Q={\rm const}\times
    \left|\,\mbox{\boldmath$I$}\,\right|^{-1/2}.  \label{prefactor}
    \end{eqnarray}
This implies the relation (\ref{DetFstar}) \cite{det}. Assembling
(\ref{onshellexponential1})) and (\ref{prefactor}) in the expression
(\ref{I2}) for $K(n_0)$ and substituting into (\ref{nintegral}) we
see that the $Q$-term of (\ref{onshellexponential1}) cancels a
similar term in the exponential of (\ref{nintegral}) and
    \begin{eqnarray}
    &&P=\left|\,\mbox{\boldmath$I$}\,\right|^{-1/2}
    \int_{-\infty}^\infty d(n_0 T)\,
    \exp\left\{\frac12\,n_0^2\,T^2\,\left(\,\frac1{\mbox{\boldmath$I$}}
    -\frac{d^2F}{d\eta^2}\, \right)\right\}
    ={\rm const}\left|\,1-\mbox{\boldmath$I$}
    \,\frac{d^2F}{d\eta^2}\,\right|^{-1/2},
    \end{eqnarray}
which confirms the gauge-independent status of the calculational
procedure and finally proves the main result
(\ref{Y})-(\ref{prefactor0}). Also, this result remains finite for
$\mbox{\boldmath$I$}=0$, which eliminates the necessity to gauge out
the second zero mode of $\mbox{\boldmath$F$}$ mentioned
above\footnote{Note that the degeneration of $\mbox{\boldmath$I$}$
to zero implies in view of (\ref{prefactor}) that ${\rm
Det_*}{\mbox{\boldmath$F$}}=0$ in full accordance with the fact that
the starred determinant includes the vanishing eigenvalue of this
zero mode $\mbox{\boldmath$\varPsi$}(\tau)$ different from
$g(\tau)$.}.

\section{Conclusions}

Thus we have derived the closed algorithm for the one-loop
contribution to the statistical sum of a generic
time-parametrization invariant gravitational model with the
Friedman-Robertson-Walker (FRW) metric. The universality of this
algorithm follows from the fact that the local part of the effective
action (\ref{action1}) is given by a generic Lagrangian ${\cal
L}(a,a')$ only restricted by the condition that it does not contain
higher derivatives of $a(\tau)$, and the nonlocal part $F(\oint
d\tau\,N/a)$ is a generic function of the conformal invariant -- the
circumference of the periodic history measured in units of the
conformal time. A universal feature of the formalism is the fact
that the quadratic part of the action is parameterized by one
function $g(\tau)$ which is both the zero mode and the generator of
the residual conformal Killing symmetry (\ref{conformaltransf}) --
inalienable feature of any FRW metric. This situation, which is
caused by a distinguished role of this metric -- the minisuperspace
sector of the theory, results in the merger of two symmetries at the
overlap of two local groups -- diffeomorphisms and conformal
transformations. Their treatment within the gauge-fixing procedure
invokes the Batalin-Vilkovisky quantization method for systems with
linearly dependent generators \cite{BFV}. It leads to the one-loop
prefactor expressed via the restricted functional determinant of the
quantum-mechanical operator with the zero mode gauged out. For the
latter we derive in quadratures a closed algorithm in terms of the
zero mode of the above type \cite{det}. The application of this
algorithm to the concrete model -- the CFT driven cosmology
suggested in \cite{slih,why} -- will be presented in the sequel to
this paper \cite{oneloop}.

The formalism of the above type has been developed for a limited set
of saddle-point solutions having one oscillation of the cosmological
scale factor. This restriction is currently explained by the fact
that the major ingredient of our result -- the restricted functional
determinant of the operator $\mbox{\boldmath$F$}$ is known in the
form (\ref{DetFstar}) only for this simplest set of instantons --
see the derivation in the accompanying paper \cite{det}. However,
the path integral applications in cosmology driven by a conformal
field theory suggest cosmological instantons with arbitrary number
of oscillations of the cosmological scale factor, corresponding to
numerous nodes of the oscillating zero mode $g(\tau)$ \cite{slih}.
In particular, for the number of these oscillations tending to
infinity this CFT driven cosmology approaches a new quantum gravity
scale -- the maximum possible value of the cosmological constant
\cite{slih,why} -- where the physics and, in particular, the effects
of the quantum prefactor become very interesting and important. Thus
the extension of the above results to an arbitrary number of
oscillations of $a(\tau)$ becomes important, as this extension --
the subject of our further study -- might be relevant to the
cosmological constant problem.

\section*{Acknowledgements}
I am indebted to A.Yu.Kamenshchik for fruitful discussions and,
especially, for clarifying the situation with the second zero mode
of the dynamical operator. I wish to express my gratitude to G.Dvali
for hospitality at the Physics Department of the Ludwig-Maximilians
University in Munich where this work was supported in part by the
Humboldt Foundation. This work was also supported by the RFBR grant
No 11-01-00830.

\appendix
\renewcommand{\thesection}{Appendix \Alph{section}.}
\renewcommand{\theequation}{\Alph{section}.\arabic{equation}}

\section{Ward identities and gauge independence
in reducible gauge theories}

Here we prove the gauge-independence relations
(\ref{omegavariation})-(\ref{sigmavariation}) of a generic
ghost-for-ghost gauge fixing procedure. The variation of the
numerator of (\ref{omegavariation}) follows from
   \begin{eqnarray}
    \delta_\sigma\ln{\rm Det}\,
    {\cal F}^\alpha_\beta=G^\beta_\alpha\,
    \delta\sigma^\alpha_a\,\omega^a_\beta,  \label{A11}
    \end{eqnarray}
where $G^\beta_\alpha$ is the inverse of the gauge-fixed ghost
operator
   \begin{eqnarray}
    {\cal F}^\alpha_\nu\,G^\nu_\beta=\delta^\alpha_\beta.
    \end{eqnarray}
Contracting this equation with $\bar Z^a_\alpha$ and taking into
account that this is a left zero-vector of $Q^\alpha_\nu$ we have
$\overline q^a_b\,\omega^b_\nu\,G^\nu_\beta=\bar Z^a_\beta$.
Therefore we have the Ward identity for the Green's function of the
ghost operator involving the inverse of the ghost-for-ghost operator
(\ref{barq0})
    \begin{eqnarray}
    \omega^a_\beta\,G^\beta_\alpha=
    (\overline q^{-1})^a_b\,\bar Z^b_\alpha.   \label{Ward})
    \end{eqnarray}
Using this in (\ref{A11}) gives
    \begin{eqnarray}
    \delta_\sigma\ln{\rm Det}\,
    {\cal F}^\alpha_\beta=
    \delta_\sigma \overline q^b_a\,
    (\overline q^{-1})^a_b=
    \delta_\sigma\ln{\rm Det}\,\overline q^a_b,  \label{A2}
    \end{eqnarray}
which proves the $\sigma$-independence in (\ref{omegavariation}).
The proof of $\omega$-independence is similar and is based on the
Ward identity complementary to (\ref{Ward}),
$G^\beta_\alpha\,\sigma^\alpha_a=(q^{-1})^b_a\,Z^\beta_b$, which in
its turn follows from contracting the alternative equation for
$G^\alpha_\beta$, $G^\alpha_\nu\,{\cal
F}^\nu_\beta=\delta^\alpha_\beta$ with the right zero-vector
$Z^\beta_b$.

To prove (\ref{sigmavariation}) note that the redundant set of gauge
conditions in this relation can formally be rewritten in terms of
independent variables $E^A$ as linear combinations
$\chi^\alpha=\chi^\alpha_A E^A$, where the coefficients identically
satisfy the relations $\bar Z^a_\alpha\chi^\alpha_A\equiv0$ with the
left zero-eigenvalue eigenvectors of the (degenerate) ghost operator
$Q^\alpha_\beta$. Therefore, the gauge conditions (corrected by
extra ghosts) $\phi^\alpha\equiv\sigma^\alpha_a E^a+\chi^\alpha$ in
(\ref{intoverg}) are in one to one correspondence with $(E^a,E^A)$
and
    \begin{eqnarray}
    E^a=(\overline q^{-1})^a_b\bar Z^b_\alpha\phi^\alpha.
    \end{eqnarray}
The variables $E^A$ and the relevant coefficients $\chi^\alpha_A$
are independent of gauge-fixing procedure for $Q^\alpha_\beta$, i.
e. of the choice of $\sigma^\alpha_a$-parameters, and
    \begin{eqnarray}
    &&\delta_\sigma\Big(\delta\big[\,\phi^\alpha\big]\Big)=
    \delta\big[\,E^a\big]\,\delta\big[\,E^A\big]\,
    \delta_\sigma\left(\frac{\partial(\,\phi^\alpha\,)}
    {\partial(E^a,E^A)}\right)^{-1}\nonumber\\
    &&\qquad\qquad\quad=
    -\delta\big[\,\phi^\alpha\big]\,
    \delta_\sigma{\ln\rm Det}[\,\sigma^\alpha_a,\chi^\alpha_A\,]
    =-\delta\big[\,\phi^\alpha\big]\,\delta\sigma^\alpha_a\frac{\partial
    E^a}{\partial\phi^\alpha}=\delta\big[\,\phi^\alpha\big]\,\delta_\sigma\ln{\rm
    Det}\,\overline q^a_b
    \end{eqnarray}
which proves the $\sigma$-independence relation
(\ref{sigmavariation}).

\section{Delta function type redundant gauge conditions}

In practice, the reduction of $\chi^\alpha$ to a set of independent
gauge conditions is not useful, especially in our case of interest
(\ref{fullgauge}), when such a reduction becomes nonlocal in time. A
better way is an explicit integration over a subset of variables
enforced by the delta function. To see this we write down the
following integral with an arbitrary test functional
$\varPhi[\,n(\tau),E\,]$ over its functional and numerical arguments
    \begin{eqnarray}
    &&\int Dn\,dE\;
    \delta\big[\,n'(\tau)+\sigma(\tau)E\,\big]\,
    \varPhi[\,n(\tau),E\,]\nonumber\\
    &&\qquad\qquad=\varPhi\left[\,
    \frac{\delta}{i\,\delta J(\tau)},\frac\partial{i\,\partial
    j}\,\right]\,\left.\int Dn\,D\pi\,dE\;
    \exp\Big(\,iS_{\rm eff}[\,n,\pi,E;\,J,I,j\,]
    \Big)\;
    \right|_{J=I=j=0},                           \label{A1}\\
    &&\nonumber\\
    &&S_{\rm eff}[\,n,\pi,E;\,J,I,j\,]
    =\oint d\tau\,\Big[\,\pi
    (n'+\pi\sigma\,E)+Jn+I\pi\,\Big] +Ej
    \end{eqnarray}
This is the Gaussian functional integral over
$\varPhi^i=(n(\tau),\pi(\tau),E)$ with the sources
$(J(\tau),I(\tau),j)$ dual to their relevant integration variables
in the ``effective" action $S_{\rm eff}[\,\varPhi;\,J,I,j\,]$. The
Gaussian integration
    \begin{eqnarray}
    &&\int D\varPhi\;
    \exp\Big(\,iS_{\rm eff}[\,\varPhi;\,J,I,j\,]
    \Big)=\left(\,{\rm Det}\,\frac{\delta^2 S_{\rm
    eff}}
    {\delta\varPhi^i\, \delta\varPhi^k}\,\right)^{-1/2}
    \exp\Big(\,iS_{\rm eff}[\,\varPhi;\,J,I,j\,]
    \Big)\;
    \Big |_{\;\rm onshell}    \label{Gaussian}
    \end{eqnarray}
restricts the exponentiated action to its stationary point -- the
solution of variational equations including
   \begin{eqnarray}
    \frac{\delta S_{\rm eff}}{\delta n(\tau)}=-\pi'(\tau)+J(\tau)=0.
    \end{eqnarray}
In view of the periodicity of $n(\tau)$ the solution of this
equation exists only when $\oint d\tau\,J=0$, which means that the
result of the functional integration here is expected to be of the
delta-function type $\sim\delta(\oint d\tau\,J)$. A possible way to
obtain this result is to regularize the exponential so that the
Gaussian integral will get a well-defined stationary point for
arbitrary sources. We will use the regularization by a small
quadratic in $n(\tau)$ term vanishing in the limit $\varepsilon\to
0$,
    \begin{eqnarray}
    S_{\rm eff}^{(\varepsilon)}[\,n,\pi,E;\,J,I,j\,]
    =\oint d\tau\,\left(\pi
    n'+\pi\sigma\,E+Jn+I\pi\right)
    +Ej-\frac\varepsilon{2i}
    \left(\oint d\tau\,n\right)^2.
    \end{eqnarray}
The corresponding variational equations take the form
   \begin{eqnarray}
   &&n'+\sigma\,E+I=0,                    \label{vareq1}\\
    &&-\pi'+J+i\varepsilon\oint d\tau_1\,n_1=0,  \label{vareq2}\\
    &&\oint d\tau\,\pi\sigma+j=0.                \label{vareq3}
    \end{eqnarray}
Integrating the first of equations over the period of $\tau$ we
immediately find
    \begin{eqnarray}
    E=-\frac1{\overline q}\oint d\tau\,I(\tau),  \label{EofI}
    \end{eqnarray}
where $\overline q$ is defined by (\ref{barq}), and
    \begin{eqnarray}
    &&n'=-\tilde I,\,\,\,\,\,\tilde I(\tau)=
    I(\tau)-\frac{\sigma(\tau)}{\overline q}\oint
    d\tau_1\,I_1,\,\,\,\,\oint d\tau\,\tilde I\equiv 0,\\
    &&n=n(0)-\int_0^\tau d\tau_1\,\tilde I_1.
    \end{eqnarray}
The constant of integration follows from the observation that the
integral of Eq.(\ref{vareq2}) over the period of $\tau$ gives $\oint
d\tau\,J+i\varepsilon T\oint d\tau\,n=0$, and $n(\tau)$ finally
equals
    \begin{eqnarray}
    n=-\int_0^\tau d\tau_1\,\tilde I_1
    +\frac1T\oint d\tau_1\int_0^{\tau_1}
    d\tau_2\,\tilde I_2
    +\frac{i}{\varepsilon T^2}
    \oint d\tau_1\,J_1.                     \label{nofJ}
    \end{eqnarray}
Then the integration of Eq.(\ref{vareq2}) gives $\pi(\tau)$ with the
integration constant following from (\ref{vareq3})
    \begin{eqnarray}
    &&\pi=\int_0^\tau d\tau_1\,\tilde J_1
    -\frac1{\overline q}\oint d\tau_1\,\sigma_1\int_0^{\tau_1} d\tau_2\,\tilde J_2
    -\frac{j}{\overline q},\\
    &&\tilde J(\tau)=J(\tau)-\frac1{T}\oint
    d\tau_1\,J_1,\,\,\,\,\,\oint d\tau\,\tilde J\equiv 0.
    \end{eqnarray}

Thus, the regularized action has a unique well-defined stationary
configuration, and the onshell value of the action with the source
$I$ switched off, $I=0$, reads
    \begin{eqnarray}
    iS_{\rm eff}^{(\varepsilon)}[\,n,\pi,E;\,J,I,j\,]\,
    \Big|_{I=0,\,\rm onshell}
    =-\frac1{2\,\varepsilon T^2}
    \left(\oint d\tau\,J\right)^2.   \label{onshellaction}
    \end{eqnarray}
Note that it is independent of the source $j$.

The preexponential factor of (\ref{Gaussian}) parametrically depends
only on the function $\sigma(\tau)$ and the regularization parameter
$\varepsilon$. This dependence can be obtained from the variational
equation
    \begin{eqnarray}
    \delta_\sigma\ln\,{\rm Det}\,\frac{\delta^2 S_{\rm
    eff}^{(\varepsilon)}}{\delta\varPhi^i \delta\varPhi^k}=
    \left(\delta_\sigma \frac{\delta^2 S_{\rm
    eff}^{(\varepsilon)}}{\delta\varPhi^i \delta\varPhi^k}\right)
    G^{ki},
    \end{eqnarray}
where $G^{ki}=-\delta\varPhi^k/\delta J_i$ is the Green's function
of the problem (\ref{vareq1})-(\ref{vareq3}) for the variables
$\varPhi^k=(n(\tau),\pi(\tau),E)$ as functionals of their sources
$J_i=(J(\tau),I(\tau),j)$. With the only nonvanishing component of
the $\sigma$-variation -- mixed functional and partial derivative
with respect to $\pi(\tau)$ and $E$ respectively,
    \begin{eqnarray}
    \delta_\sigma\left(\frac\delta{\delta\pi(\tau)}
    \frac\partial{\partial E}S_{\rm eff}^{(\varepsilon)}\right)=
    \delta\sigma(\tau),
    \end{eqnarray}
we have
    \begin{eqnarray}
    \delta_\sigma\ln\,{\rm Det}\,\frac{\delta^2 S_{\rm
    eff}^{(\varepsilon)}}{\delta\varPhi^i \delta\varPhi^k}=
    -2\oint d\tau\,\delta_\sigma\!\left(\frac\delta{\delta\pi(\tau)}
    \frac\partial{\partial E}S_{\rm eff}^{(\varepsilon)}\right)
    \frac{\delta E}{\delta I(\tau)}
    =\frac2{\overline q}\oint
    d\tau\,\delta\sigma
    =2\delta_\sigma\ln\overline q,      \label{sigmavariation1}
    \end{eqnarray}
where we have used Eq.(\ref{EofI}).

Similarly $\delta_\varepsilon(\delta^2S_{\rm
eff}^{(\varepsilon)}/\delta n_1\delta n_2)=i\delta\varepsilon$ and
from (\ref{nofJ}) $\delta n_2/\delta J_1=i/\varepsilon T^2$, so that
    \begin{eqnarray}
    \delta_\varepsilon\ln\,{\rm Det}\,\frac{\delta^2 S_{\rm
    eff}^{(\varepsilon)}}{\delta\varPhi^i \delta\varPhi^k}
    =-\oint d\tau_1\,d\tau_2,\delta_\varepsilon\!
    \left(\frac{\delta^2S_{\rm eff}^{(\varepsilon)}}
    {\delta n_1\delta n_2}\right)
    \frac{\delta n_2}{\delta
    J_1}=\delta_\varepsilon\ln\varepsilon.  \label{epsilonvariation}
    \end{eqnarray}

Collecting (\ref{onshellaction}), (\ref{sigmavariation1}) and
(\ref{epsilonvariation}) we obtain in the limit of $\varepsilon\to
0$ the anticipated delta function
    \begin{eqnarray}
    &&\left.\int Dn\,D\pi\,dE\;
    \exp\Big(\,iS_{\rm eff}^{(\varepsilon)}
    [\,n,\pi,E;\,J,I,j\,]\Big)\;\right|_{I=0}\nonumber\\
    &&\qquad\qquad\qquad=
    {\rm const}\times\frac1{\overline q\sqrt\varepsilon}
    \exp\left[-\frac1{2\varepsilon T^2}\left(\oint
    d\tau\,J\right)^2\right]={\rm const}\,\frac{T}{\overline q}\,
    \delta\!\left(\oint d\tau\,J\right)
    \end{eqnarray}
Note that the result is independent of the source $j$ dual to the
extra ghost $E$. Using this result in (\ref{A1})
    \begin{eqnarray}
    &&\int Dn\,dE\;
    \delta\big[\,n'(\tau)+\sigma(\tau)E\,\big]\,
    \varPhi[\,n(\tau),E\,]\nonumber\\
    &&\qquad\qquad={\rm const}\,\frac{T}{\overline q}\,\varPhi\left[\,
    \frac{\delta}{i\,\delta J(\tau)},\frac\partial{i\,\partial
    j}\,\right]\,\left.\int_{-\infty}^{\infty} dn_0\,
    \exp\left(in_0\oint d\tau\,J\right)\;\right|_{J=j=0}\nonumber\\
    &&\qquad\qquad={\rm const}\,\frac{T}{\overline q}\,
    \int_{-\infty}^{\infty} dn_0\,
    \varPhi[\,n_0,0\,]
    \end{eqnarray}
we finally prove (\ref{integralovernE}).

\end{document}